\documentclass[12pt,preprint]{aastex}
\usepackage{natbib,times}

\shorttitle{} \shortauthors{}
\begin{document}

\title{Can FSRQs produce the IceCube detected diffuse neutrino emission?}

\author{Bin Wang$^{1}$ and Zhuo Li$^{1,2}$}
\affil{$^{1}$Department of Astronomy, School of Physics, Peking University, Beijing 100871, China; \texttt{wang\_b@pku.edu.cn; zhuo.li@pku.edu.cn}\\
$^{2}$Kavli Institute for Astronomy and Astrophysics, Peking
University, Beijing 100871, China}

\begin{abstract}
IceCube has reported the detection of a diffuse TeV-PeV neutrino
emission, for which the flat spectrum radio quasars (FSRQs) have
been proposed to be the candidate sources. Here we assume that the
neutrino flux from FSRQs is proportional to their gamma-ray ones,
and obtain the gamma-ray/neutrino flux ratio by the diffuse
gamma-ray flux from \textit{Fermi}-LAT measurement of FSRQs and the
diffuse neutrino flux detected by IceCube. We apply this ratio to
individual FSRQs and hence predict their neutrino flux. We find that
a large fraction of candidate FSRQs from the northern sky in the
IceCube point source search has predicted neutrino flux above the
IceCube upper limit; and for the sample of stacking search for
neutrinos by IceCube, the predicted stacked flux is even larger than
the upper limit of stacked flux by orders of magnitude. Therefore
the IceCube limit from stacking searches, combined with the
\textit{Fermi}-LAT observations, already rejects FSRQs as the main
sources of IceCube-detected diffuse neutrinos: FSRQs can only
account for $\lesssim10\%$ ($\lesssim4\%$) of the IceCube-detected
diffuse neutrino flux, according to the stacking searches from the
whole (northern) sky. The derived neutrino/gamma-ray flux ratio is
$\ll1$, also implying that the gamma-ray emission from FSRQs cannot
be produced by the secondary leptons and photons from the pion
production processes. The caveat in the assumptions is discussed.
\end{abstract}


\section{Introduction}
The detection of extraterrestrial diffuse TeV-PeV neutrino emission
has been reported recently by IceCube collaboration
\citep{2PeV,ic13,ic14} with an excess to the atmospheric background
up to 5.7$\sigma$ \citep{ic14}. The single flavor neutrino spectrum
can be fitted with a flat form \citep{ic14}
\begin{equation}
  E_\nu^2 \Phi_{\nu,\rm IC}= (0.95\pm0.3)\times 10^{-8} \rm GeV cm^{-2}s^{-1}sr^{-1}
\end{equation}
from tens TeV up to 2 PeV, assuming flavor ratio of 1:1:1. The lack
of neutrino events above 2 PeV may suggest a cutoff at the neutrino
spectrum. The sky distribution of the neutrino events is consistent
with being isotropic \citep{ic14} and hence of extragalactic origin,
but the low statistics so far cannot make any solid conclusion.
There are many models proposed to explain these diffuse neutrinos,
e.g., produced by Galactic cosmic ray (CR) propagation or Galactic
point sources
\citep{gupta13,neronov13,ahlers14,razzaque13-1,razzaque13-2,joshi14,guo13,taylor14},
by extragalactic sources such as gamma-ray bursts (GRBs)
\citep{waxman97,murase13a,cholis13,liu13a}, active galactic nuclei
(AGN) jets \citep{murase14,dermer14,BLLac2,BLLac1} or cores
\citep{stecker13}, star forming galaxies
\citep{murase13b,liu13b,he13,anchordoqui14,tamborra14} and
starbutrst galaxies \citep{loeb06,wang14}, and by cosmogenic
neutrinos from ultrahigh energy CR interactions with cosmic
microwave background (CMB) \citep{kalashev13,laha13,roulet13}. By
using the Fermi-LAT observations and considering the gamma-ray and
neutrino flux correlation, \cite{wang14} suggested that only
starburst galaxies and AGN jets in the above mentioned scenarios may
produce the neutrino flux detected by IceCube. We further
investigate the AGN jet origin in this paper.

AGNs have long been considered as ultrahigh energy CR sources, and
hence producing neutrinos through photo-pion processes. Ref
\cite{murase14,dermer14} suggested that CRs accelerated in the inner
jets of AGNs, especially flat spectrum radio quasars (FSRQs),
interact with photons from the broad-line region (BLR), can produce
neutrinos that may account for the IceCube detected PeV neutrino
flux. The \textit{Fermi} Large Area Telescope (LAT) already provides
us a deep survey of the sky from about $30$ MeV to several hundred
GeV. Based on the 2-year observation of \textit{Fermi}-LAT, the
FSRQs' luminosity function (LF) and redshift distribution are well
determined \citep{fsrq}. Assuming a correlation between the
gamma-ray and neutrino fluxes from FSRQs, we constrain the neutrino
emission by FSRQs using gamma-ray flux from \textit{Fermi}-LAT
observations, in order to examine the FSRQ scenario for IceCube
detected neutrinos. In Section 2 we describe the assumption and the
method of the analysis, Section 3 shows the data and results, and
conclusion and discussion are presented in Section 4.

\section{Assumption and method}
We assume that (1) \emph{the gamma-ray and neutrino fluxes from
FSRQs are proportional to each other;} and (2) FSRQs can account for
the diffuse neutrino flux detected by IceCube. The following
analyses are all carried out under these two assumptions.

There are leptonic and hadronic models for blazar gamma-ray emission
\citep{blazar-gamma}. In the leptonic model the FSRQ gamma-ray
emission is produced by electrons accelerated in the inner jet
upscattering background soft photons from self produced synchrotron
radiation or the accretion disk and BLR. On the other hand, in the
hadronic model the FSRQ gamma-ray emission is mainly produced by CR
proton synchrotron radiation\footnote{We can rule out that in the
hadronic model the gamma-ray emission from FSRQs can be produced by
secondary leptons and photons from photopoin process, given the
result of eq. (4) below. See the discussion in the last section.}.
In both cases assumption (1) is reasonable if the energy ratio
between the jet accelerated electrons and CRs is constant. Moreover,
assumption (1) should be more reasonable and valid in a statistic
sense, although it may not tightly hold for individual FSRQs.

The FSRQs are bright sources in the gamma-ray sky. The luminosity
function (LF) and redshift evolution of FSRQs have well been
developed by \cite{fsrq} based on the 2 years survey by
\textit{Fermi}-LAT. The all sky (i.e., $4\pi$ integrated) diffuse
gamma-ray emission produced by FSRQs is determined to be
\begin{equation}
  J_{\gamma}=1.45\times10^{-5}\rm GeVcm^{-2}s^{-1}
\end{equation}
in the energy range of 0.1-100 GeV (see more details for the
derivation in the appendix).

According to the flat spectrum IceCube detected (eq. 1), the diffuse
single-flavor neutrino flux in the energy range of 20 TeV-2 PeV is
\begin{equation}
\label{eq:4} J_{\nu}=4\pi\int_{\rm 20 TeV}^{\rm 2
PeV}E_\nu\Phi_{\nu,\rm IC}dE_\nu=4\pi\ln(10^2)E_\nu^2 \Phi_{\nu,\rm
IC}=5.5\times 10^{-7} \rm GeV cm^{-2}s^{-1}.
\end{equation}
If FSRQs can account for the IceCube detected flux, the ratio of
neutrino (20 TeV-2 PeV) to $\gamma$-ray (0.1-100 GeV) flux for FSRQs
is then
\begin{equation}
\label{eq:5} J_{\nu}/J_{\gamma}=3.79\times10^{-2}.
\end{equation}
Applying this ratio to individual FSRQs with gamma-ray flux measured
we can estimate their neutrino flux.

Analyzing the 4-year data, the IceCube presented results of
searching for astrophysical neutrinos from candidate sources
\citep{pointnu}. The non-detection only gives upper limits to the
candidate sources based on the IceCube sensitivity. For an $E^{-2}$
spectrum, the median sensitivity is $ \sim 10^{-12} \rm
TeV^{-1}cm^{-2}s^{-1}$ in the northern sky for energies of
1TeV-1PeV, and $ \sim 10^{-11} \rm TeV^{-1}cm^{-2}s^{-1}$ in the
southern sky for energies of 0.1-100PeV \citep{pointnu}. Thus we pay
more attention to the candidate objects in the northern sky since
they are more strongly constrained by IceCube observations. We will
compare the predicted neutrino fluxes to the limits given by IceCube
for individual and stacked FSRQs, thus examine the FSRQ origin of
the diffuse neutrino emission.

\section{Data and results}
\subsection{Candidate sources}
IceCube has searched for neutrino emission from 44 candidate
sources, among which there are 11 FSRQs, 5 in the northern sky and 6
in the southern sky \citep{pointnu}. The non-detection by IceCube
puts upper limits of neutrino flux to these sources. With the
gamma-ray/neutrino flux ratio of Eq.~\ref{eq:5}, we predicted the
neutrino flux from them using gamma-ray flux for individual FSRQs
from the \textit{Fermi}-LAT 2-year point source catalog
(2FGL)\footnote{\url{http://fermi.gsfc.nasa.gov/ssc/data/access/lat/2yr\_catalog}}.
The results are presented in Table 1, and the comparison between
predicted flux and the IceCube limit is shown in Fig 1.

The sensitivity of IceCube for point sources depends strongly on the
source declination, as shown in Fig. 11 in \cite{pointnu} (also
shown in Fig 2 here). The typical sensitivity in the southern sky,
$\sim 10^{-11}\rm TeV^{-1}cm^{-2}s^{-1}$, is much larger than the
predicted neutrino fluxes. However, the sensitivities in the
northern sky is much higher, and 3 of the 5 northern sources are
with predicted fluxes above the upper limits. In details, the
predicted flux of the source 3C 454.3 is larger than the upper limit
by a factor of 7.3, PKS 1506+106 by 1.4, and 3C 273 by 2.0.
\begin{figure}
\includegraphics[width=\columnwidth]{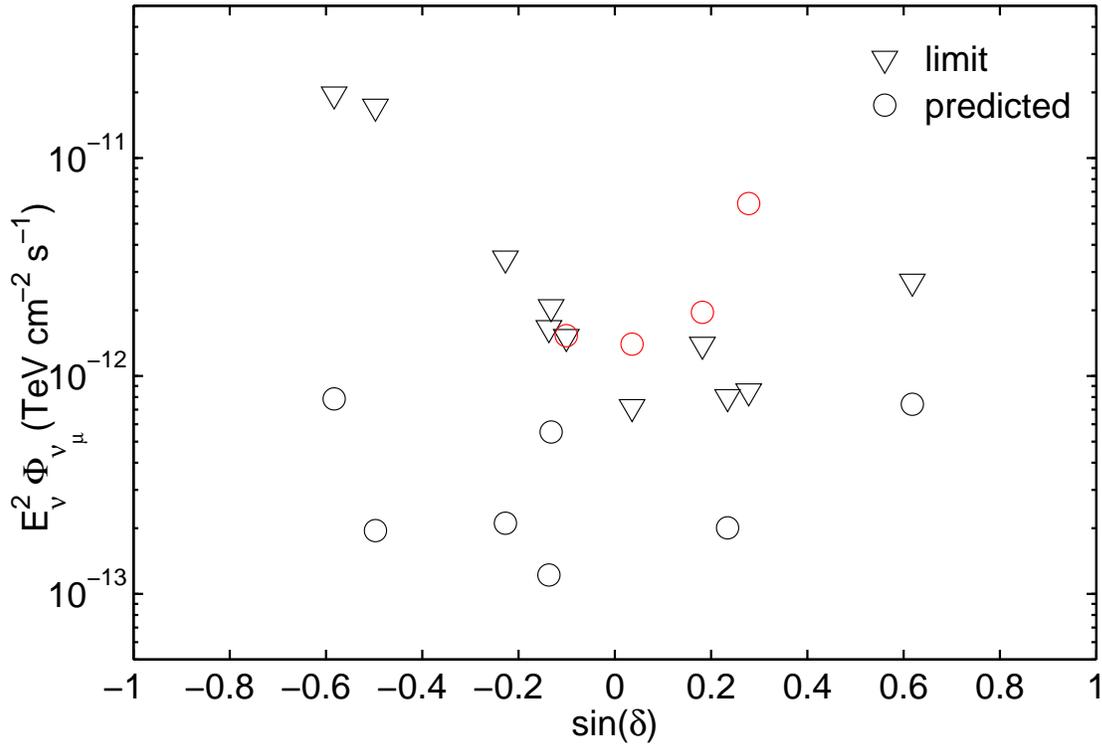}
\centering \caption{The predicted fluxes and the upper limits for
candidate FSRQs in Table 1 as function of $\sin(\delta)$, with
$\delta$ the FSRQ declination. The \emph{circles} show the predicted
single flavor neutrino flux of individual FSRQs, and the
\emph{inverse triangles} the upper limits by IceCube of 4-year
running. Those marked in \emph{red} are with predicted flux larger
than the IceCube limit: 3C 279, 3C 273, PKS 1502+106, and 3C 454.3
(from left to right).}
  \label{fig:1}
\end{figure}

\begin{table}
  \caption[]{Gamma-ray and neutrino fluxes from candidate FSRQs.}
  \label{Tab:limit}
  \begin{center}\begin{tabular}{cccccc}
  \hline\noalign{\smallskip}
FSRQs &  r.a.  & dec   & $F_{\gamma}$  & $E_\nu^2\Phi_{\nu_{\mu},\rm pred}$ & $E_\nu^2\Phi_{\nu_{\mu},\rm lim}$  \\
  \hline\noalign{\smallskip}
Northern Sky\\
4C 38.41 & 248.81 & 38.17 & 1.44 & 0.74 & 2.71  \\
\textbf{3C 454.3} & 343.50 & 16.15 & 12.02 & 6.18 & 0.85 \\
PKS 0528+134 & 82.71 & 13.55 & 0.39 & 0.20 & 0.80  \\
\textbf{PKS 1502+106} & 226.10 & 10.49 & 3.82 & 1.96 & 1.39  \\
\textbf{3C 273} & 187.28 & 2.04 & 2.72 & 1.40 & 0.72 \\
  \hline\noalign{\smallskip}
Southern Sky\\
\textbf{3C 279} & 194.04 & -5.79 & 2.98 & 1.53 & 1.51 \\
QSO 2022-077 & 306.42 & -7.61 & 1.08 & 0.55 & 2.07  \\
PKS 1406-076 & 212.24 & -7.87 & 0.24 & 0.12 & 1.66  \\
QSO 1730-130 & 263.28 & -13.13 & 0.41 & 0.21 & 3.46 \\
PKS 1622-297 & 246.53 & -29.81 & 0.30 & 0.20 & 17.2 \\
PKS 1454-354 & 224.36 & -35.67 & 1.21 & 0.79 & 19.6 \\
  \noalign{\smallskip}\hline

  \end{tabular}\end{center}
Note - The columns from left to right show: the FSRQ names; the
right ascension and declination of the sources; the gamma-ray flux
measured by \textit{Fermi}-LAT in the 0.1-100GeV energy range (in
unit of $10^{-10}\rm erg cm^{-2} s^{-1}$); the predicted
$\nu_\mu+\bar{\nu}_\mu$ flux in unit of $10^{-12} \rm TeV cm^{-2}
s^{-1}$, assuming $E^{-2}$ flux normalization and even flavor ratio
of 1:1:1; and the upper limit of $\nu_\mu+\bar{\nu}_\mu$ flux (in
unit of $10^{-12} \rm TeV cm^{-2} s^{-1}$) in the 90\% confidence
level from IceCube, assuming $E^{-2}$ flux normalization. The source
names in bold face mark those FSRQs with predicted flux violating
the upper limit.
\end{table}

\subsection{Stacking sample}
Since it is usually null result for detection of individual sources,
IceCube also carried out stacking searches for different classes of
objects \citep{pointnu}. For FSRQs, they select 33 sources based on
the measured \textit{Fermi}-LAT gamma-ray flux, assuming prevalence
of photo-hadronic neutrino production. The upper limit in the 90\%
confidence level, for a $E^{-2}$ flux normalization of
$\nu_\mu+\bar{\nu}_\mu$ flux, is $3.46\times 10^{-12} \rm
TeV\,cm^{-2}s^{-1}$ for the total flux of the whole sample in the
stacking analysis. So the limit for each FSRQ on average obtained
from stacking is smaller by 33,
\begin{equation}
  E_\nu^2\Phi_{\nu_\mu,\rm lim}=1.05\times 10^{-13} \rm
TeV\,cm^{-2}s^{-1}.
\end{equation}
This is already smaller than all the predicted flux for individual
FSRQs in the sample list (see Table 2).

Since the detection of muon and anti-muon neutrinos by IceCube
depends sensitively on the declination of the source, we may focus
more on the directions of the sky with higher sensitivity,
especially the northern sky. The FSRQ sample in the stacking search
by IceCube includes 19 in the northern sky and 14 in the southern
sky. We here derive the upper limit for the 19 northern ones.

As the upper limits are basically given by the sensitivity, and the
sensitivity as function of declination ($\delta$) has been given by
IceCube (Fig 11 in \citep{pointnu}), we may estimate the upper limit
of individual sources by the sensitivity at their declinations. The
ratio of the upper limit for stacking 19 northern FSRQs,
$f_{\nu_\mu,\rm lim}^{\rm north}$, to that for the whole-sky 33
ones, $f_{\nu_\mu,\rm lim}^{\rm total}=33\Phi_{\nu_\mu,\rm lim}$, is
\begin{equation}
\label{eq:6} \eta=\frac{f_{\nu_\mu,\rm lim}^{\rm
north}}{f_{\nu_\mu,\rm lim}^{\rm
total}}=\frac{\sum\limits_{i=1}^{19} \Phi_{\nu_\mu,\rm
sens}(\delta_i)}{\sum\limits_{j=1}^{33}\Phi_{\nu_\mu,\rm
sens}(\delta_j)},
\end{equation}
where $\Phi_{\nu_\mu,\rm sens}(\delta)$ is the sensitivity at
declination $\delta$, and $\delta_i$ ($\delta_j$) is the declination
of No. $i$ ($j$) source. The average upper limit for the 19
northern-sky FSRQs, $\Phi_{\nu_\mu,\rm lim}^{\rm
north}=f_{\nu_\mu,\rm lim}^{\rm north}/19$, is then
\begin{equation}
\label{eq:8} \Phi_{\nu_\mu,\rm lim}^{\rm north}
=\frac{33}{19}\eta\Phi_{\nu_\mu,\rm lim}.
 \end{equation}
Here $\eta$ can be evaluated by the 4-year sensitivity of IceCube in
90\% confidence level (Fig.2), $\eta=0.19$ (see more details in the
appendix). Thus the upper limit of the $\nu_\mu+\bar{\nu}_\mu$ flux
per source from the northern sky is
\begin{equation}
  E_{\nu}^2\Phi_{\nu_\mu,\rm lim}^{\rm north}=3.45\times 10^{-14} \rm TeV
cm^{-2}s^{-1},
\end{equation}
which is about 3 times more stringent than the limit from the whole
sample of FSRQs (eq 5).

By the method introduced in Section 2, the predicted
$\nu_\mu+\bar{\nu}_\mu$ flux per source (i.e., the total flux
divided by the source number) for the whole sample is
\begin{equation}
  E_\nu^2\Phi_{\nu_\mu,\rm pred}=1.17\times 10^{-12} \rm TeV
cm^{-2}s^{-1}.
\end{equation}
Note, compared with the limit of eq. (5), the prediction is about 11
times larger. As for the 19 northern-sky FSRQs, the predicted
average flux after stacking is
\begin{equation}
 E_\nu^2\Phi_{\nu_\mu,\rm pred}^{\rm
north}=0.93\times 10^{-12} \rm TeV cm^{-2}s^{-1}.
\end{equation}
Thus the predicted flux is even 27 times larger than the limit for
the northern sky FSRQs (eq. 8). We show in Fig 2 the predicted
fluxes of the 33 FSRQs, as well as the predicted stacked flux and
the upper limit for the northern-source stacking.

\begin{figure}
\includegraphics[width=\columnwidth]{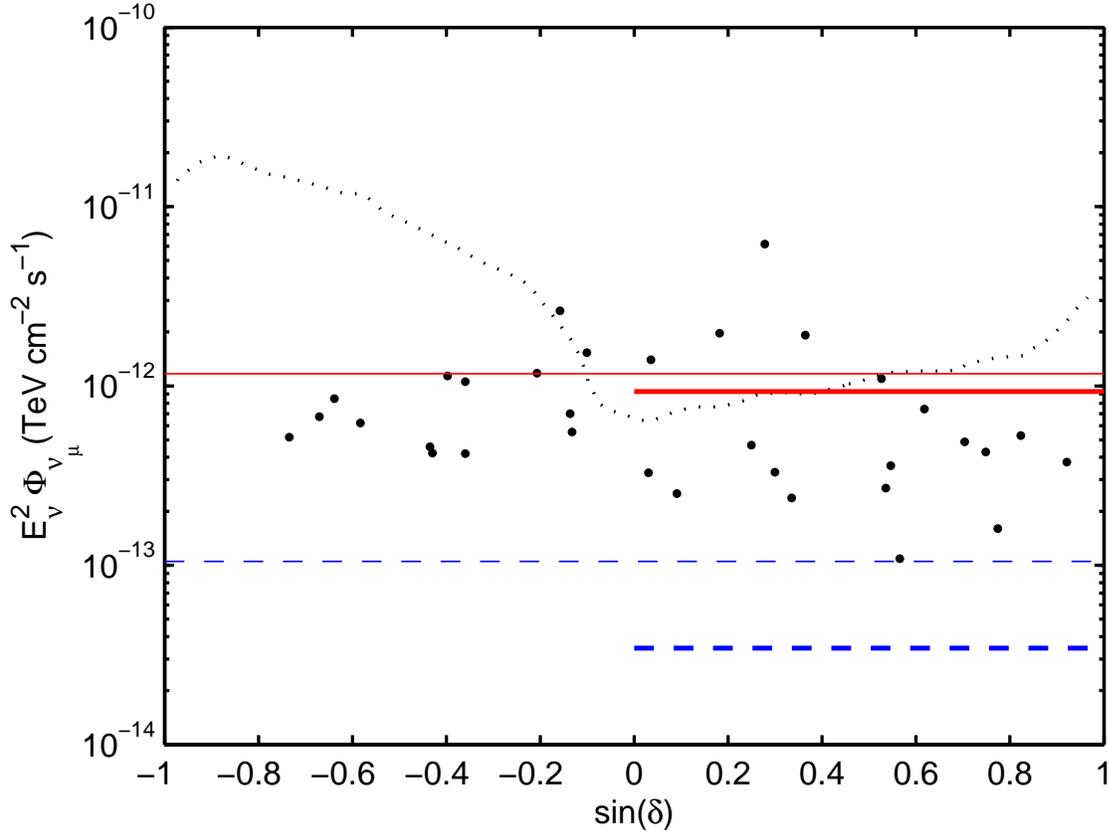}
\centering \caption{The predicted fluxes for FSRQs in the stacked
sample (Table 2; \emph{dots}) as function of $\sin(\delta)$, with
$\delta$ the declination. The \emph{red solid} lines mark the
predicted single-flavor neutrino fluxes, whereas the \emph{blue
dashed} lines are the 4-year IceCube upper limits. The \emph{thin}
(\emph{thick}) lines are for the stacked sample from the whole
(northern) sky. Also shown is the IceCube sensitivity of 4 years of
running time (\emph{black dotted} line).}
  \label{fig:2}
\end{figure}

\section{Conclusion and Discussion}
We derive the muon neutrino (and anti-muon neutrino) flux of FSRQs
under the assumptions that there is a connection between the
gamma-ray and neutrino fluxes, and that FSRQs can account for the
diffuse neutrino flux detected by IceCube. By comparing the derived
flux and the limit for FSRQs from 4 years of running time of
IceCube, we find that (i) a significant fraction of the candidate
FSRQs, 3 among 5, has derived flux larger than the upper limit, in
particular, 3C 454.3 is predicted to be exceeding the IceCube upper
limit by a factor of 7; and (ii) in the sample selected for IceCube
stacking search, the predicted flux is more than one order of
magnitude larger than the upper limit for the whole sample, and even
27 times larger for the northern sample which has stringent limit by
IceCube.

The confliction between the prediction and the observation implies
that at least one of the assumptions is wrong. If the first
assumption is valid, the second one will not hold, i.e., FSRQs are
not the main sources of the diffuse neutrino emission. By the
contrast between the predicted flux and the upper limit for the
stacked sample, FSRQs produce at most $10\%$ of the IceCube-detected
diffuse neutrino flux. Furthermore, according to the constraint from
the northern-sky sample FSRQs can only produce a diffuse neutrino
flux $<3.7\%$ of the IceCube-detected one.

As the sensitivity improves with running time (roughly
$\propto1/\sqrt{t}$), the IceCube observation in coming years will
make more stringent constraint on the FSRQ origin of diffuse
neutrino emission. In particular, the FSRQs in the sky region that
IceCube has better sensitivity are favored to be selected to
constrain their neutrino flux, and hence make more stringent
constraint on FSRQ neutrino flux, e.g., those sources in the region
of $\sin\delta\sim-0.1-0.9$ according to the IceCube sensitivity
dependence of declination (Fig 2).

It should be noted that the sample of 33 FSRQs selected by IceCube
collaboration in the stacking search is selected with the weight of
the measured gamma-ray flux by \textit{Fermi}-LAT. This approach is
consistent with our assumption that the gamma-ray and neutrino
fluxes are well correlated. So we can use this sample for our
purpose. Moreover, we note that if for some reason the faint FSRQs
produce larger neutrino emission than bright ones, then our method
and conclusion are not valid. However, since the contrast between
the prediction and the limit is by orders of magnitude, we may have
strong confidence to reject the FSRQ origin scenario for diffuse
neutrino emission.

An implication from eq (4) is that the FSRQ gamma-ray emission
cannot be produced by the secondary leptons and photons in the pion
production processes. The energy carried by neutrinos and secondary
leptons plus photons are comparable, thus one expects comparable
neutrino and gamma-ray fluxes from pion production processes,
$J_\gamma\sim J_\nu$, in contrast with eq (4). $J_\gamma\gg J_\nu$
(eq 4) can only appear in unreal cases where pions loose energy
rapidly by radiation before decay into neutrinos.

\section*{Acknowledgments}

We thank J. Feintzeig for providing the list of the FSRQ sample in
the stacking search, K. Schatto and M. Ajello for helpful
discussion, and the referee for very critical comments. This work is
partly supported by NSFC (11273005), SRFDP (20120001110064), and the
973 Program (2014CB845800).

\appendix
\section{Calculation of diffuse gamma-ray flux from FSRQs}
The all-sky integrated intensity of the diffuse gamma-ray emission
contributed by FSRQs can be derived as
\begin{equation}
\label{eq:int} J_{\gamma}=\int_{z_{min}}^{z_{max}} dz
\int^{\Gamma_{max}}_{\Gamma_{min}} d\Gamma
\int^{L_{{\gamma},max}}_{L_{{\gamma},min}} dL_{\gamma}
F_{\gamma}(L_{\gamma},z)\frac{d^3 N}{dL_{\gamma}dzd\Gamma},
\end{equation}
where $z$ is the redshift, $\Gamma$ is the source photon index,
$L_\gamma$ is the rest frame 0.1-100 GeV $\gamma$-ray luminosity,
and $F_\gamma$ is the gamma-ray flux,
\begin{equation}
\label{eq:10} F_\gamma =\frac{L_\gamma}{4\pi D_L^2 (z)},
\end{equation}
with $D_L(z)$ the luminosity distance of the source. Following
\cite{fsrq} the distribution of FSRQs can be written as
\begin{equation}
\label{eq:11} \frac{d^3 N}{dL_{\gamma} dz d\Gamma}=
\frac{d^2N}{dL_{\gamma} dV}\times \frac{dN}{d\Gamma} \times
\frac{dV}{dz} = \Phi(L_{\gamma},z) \times \frac{dN}{d\Gamma} \times
\frac{dV}{dz},
\end{equation}
where $\Phi(L_{\gamma},z)$ is the LF, $dN/d\Gamma$ is the photon
index distribution, assumed to be a Gaussian distribution with mean
$\mu$ and dispersion $\sigma$ and independent of $z$,
 \begin{equation}
\frac{dN}{d\Gamma}=e^{-\frac{(\Gamma-\mu)^2}{2\sigma^2}},
\end{equation}
and $dV/dz=4\pi D_c^2 (z)dD_c(z)/dz$ is the comoving volume element
per unit redshift, where $D_c(z)$ is the comoving distance. A
standard concordance cosmology is assumed, with $H_0=71 \rm km
s^{-1}Mpc^{-1}$ and $\Omega_M = 1- \Omega_{\Lambda}=0.27$, thus
\begin{equation}
D_c(z)=\frac{c}{H_0}\int^z_0 \frac{1}{\sqrt{(\Omega_M (1+z_1)^3 +
\Omega_\Lambda)}}dz_1.
\end{equation}
and $D_L(z)=(1+z)D_c(z)$.

For the LF, we adopt the LDDE model suggested in \cite{fsrq},
\begin{equation}
\Phi(L_{\gamma},z) = \Phi(L_{\gamma}) \times e(z,L_{\gamma})
\end{equation}
where
\begin{equation}
\Phi(L_{\gamma}) =
\frac{dN}{dL_{\gamma}}=\frac{A}{\ln(10)L_{\gamma}}
\left[\left(\frac{L_{\gamma}}{L_{*}}\right)^{\gamma_1}+
\left(\frac{L_{\gamma}}{L_{*}}\right)^{\gamma2} \right]^{-1},
\label{eq:ple1}
\end{equation}

\begin{equation}
e(z,L_{\gamma})= \left[
\left( \frac{1+z}{1+z_c(L_{\gamma})}\right)^{p1} + \left(
\frac{1+z}{1+z_c(L_{\gamma})}\right)^{p2}
 \right]^{-1},
\label{eq:evol}
\end{equation}
and
\begin{equation}
z_c(L_{\gamma})= z_c^*\cdot (L_{\gamma}/10^{48})^{\alpha}.
\label{eq:zpeak}
\end{equation}
Here $L_\gamma$ and $L_{*}$ are in unit of $10^{48}\rm erg\,s^{-1}$.

The limits of the integration in Eq.~\ref{eq:int} are
$L_{\gamma,\min}=10^{44}$\,erg s$^{-1}$,
$L_{\gamma,\max}=10^{52}$\,erg s$^{-1}$,
 $z_{\min}=0$, $z_{\max}=6$, $\Gamma_{\min}=1.8$ and $\Gamma_{\max}=3.0$.
The parameters in the LDDE model have been determined by the fitting
of the \textit{Fermi}-LAT data,
 $A=3.06(\pm 0.23)\times 10^{-9}\rm Mpc^{-3} erg^{-1}s$,
 $\gamma_1=0.21\pm 0.12$, $\gamma_2=1.58\pm 0.27$,
 $L_{*}=0.84\pm0.49$, $z_c^{*}=1.47\pm0.16$, $\alpha=0.21\pm0.03$,
 $p_1=-7.35\pm1.74$, $p_2=6.51\pm1.97$,
 $\mu=2.44\pm0.01$, and $\sigma=0.18\pm0.01$. We calculate the
 all-sky diffuse gamma-ray flux from FSRQs using these parameter values,
 and give $J_\gamma=1.45\times10^{-5}\rm GeVcm^{-2}s^{-1}$.

\section{The FSRQs in the stacking search of IceCube}
We show the sample of the stacking search in Table 2, with the
sensitivity of IceCube of the relevant declinations shown, as well
as the predicted $\nu_\mu+\bar{\nu}_\mu$ fluxes. Summing up the
sensitivity for the northern objects only and for the whole sample
gives
\begin{equation}
\eta=\frac{\sum\limits_{i=1}^{19} \Phi_{\nu_\mu,\rm
sens}(\delta_i)}{\sum\limits_{j=1}^{33}\Phi_{\nu_\mu,\rm
sens}(\delta_j)}=0.19.
\end{equation}

\begin{footnotesize}
\begin{table}
\caption{The 33 FSRQs in the IceCube stacking search}
\begin{center}
\begin{tabular}{cccc}
 \hline
FSRQs & $\sin(\delta)$ & $E_\nu^2\Phi_{\nu_\mu,\rm pred}$ & $E_\nu^2\Phi_{\nu_\mu,\rm sens}$ \\
 \hline
Northern Sky\\
2FGL J1849.4+6706 & 0.92 & 0.38 & 2.30\\
2FGL J0957.7+5522 & 0.82 & 0.53 & 1.49\\
2FGL J0654.5+5043 & 0.77 & 0.16 & 1.41\\
2FGL J1312.8+4828 & 0.75 & 0.43 & 1.40\\
2FGL J0920.9+4441 & 0.70 & 0.49 & 1.33\\
2FGL J1635.2+3810 & 0.62 & 0.74 & 1.23\\
2FGL J0043.7+3426 & 0.57 & 0.11 & 1.20\\
2FGL J0719.3+3306 & 0.55 & 0.36 & 1.20\\
2FGL J1310.6+3222 & 0.54 & 0.27 & 1.18\\
2FGL J1522.1+3144 & 0.53 & 1.10 & 1.17\\
2FGL J1224.9+2122 & 0.36 & 1.92 & 0.90\\
2FGL J0714.0+1933 & 0.33 & 0.24 & 0.93\\
2FGL J2203.4+1726 & 0.30 & 0.33 & 0.92\\
2FGL J2253.9+1609 & 0.28 & 6.18 & 0.89\\
2FGL J0725.3+1426 & 0.25 & 0.47 & 0.87\\
2FGL J1504.3+1029 & 0.18 & 1.96 & 0.77\\
2FGL J1016.0+0513 & 0.09 & 0.25 & 0.73\\
2FGL J1229.1+0202 & 0.04 & 1.40 & 0.64\\
2FGL J0217.9+0143 & 0.03 & 0.33 & 0.63\\
 \hline
Southern Sky\\
2FGL J1256.1-0547 & -0.10 & 1.53 & 1.17\\
2FGL J2025.6-0736 & -0.13 & 0.55 & 1.82\\
2FGL J0808.2-0750 & -0.14 & 0.70 & 1.83\\
2FGL J1512.8-0906 & -0.16 & 2.62 & 2.19\\
2FGL J0730.2-1141 & -0.20 & 1.17 & 3.16\\
2FGL J1833.6-2104 & -0.36 & 1.05 & 5.53\\
2FGL J1923.5-2105 & -0.36 & 0.42 & 5.53\\
2FGL J0457.0-2325 & -0.40 & 1.13 & 6.28\\
2FGL J1625.7-2526 & -0.43 & 0.42 & 6.96\\
2FGL J1246.7-2546 & -0.44 & 0.46 & 7.07\\
2FGL J1457.4-3540 & -0.58 & 0.62 & 11.64\\
2FGL J1802.6-3940 & -0.64 & 0.85 & 12.20\\
2FGL J1428.0-4206 & -0.67 & 0.67 & 13.31\\
2FGL J2056.2-4715 & -0.73 & 0.52 & 14.41\\
\hline
\end{tabular}
\end{center}
Note - The last two columns are the predicted
$\nu_\mu+\bar{\nu}_\mu$ flux and the 90\% confidence level
sensitivity (for muon and anti-muon neutrinos) of the 4 years of
IceCube running time at the declination of the relevant source. Both
are in unit of $10^{-12} \rm TeV cm^{-2} s^{-1}$.
\end{table}
\end{footnotesize}


\begin{thebibliography}{}

\bibitem[Aartsen et al.(2013)]{2PeV} Aartsen, M.~G., Abbasi, R., Abdou, Y., et al.\ 2013, \prl, 111, 021103
\bibitem[Aartsen et al.(2014a)]{ic14} Aartsen, M.~G., Ackermann, M., Adams, J., et al.\ 2014a, \prl, 113, 101101
\bibitem[Aartsen et al.(2014c)]{pointnu} Aartsen, M.~G., Ackermann, M., Adams, J., et al.\ 2014, \apj, 796, 2
\bibitem[Ahlers and Murase(2014)]{ahlers14} Ahlers, M., \& Murase, K.\ 2014, \prd, 90, 023010
\bibitem[Ajello et al. (2012)] {fsrq} Ajello, M., Shaw, M. S., Romani, R. W., et al.\ 2012, \apj, 751, 108
\bibitem[Anchordoqui et al.(2014a)]{anchordoqui14} Anchordoqui, L.~A., Goldberg, H., Lynch, M.~H., et al.\ 2014, \prd, 89, 083003
\bibitem[B{\"o}ttcher et al.(2013)]{blazar-gamma} B{\"o}ttcher, M., Reimer, A., Sweeney, K., \& Prakash, A.\ 2013, \apj, 768, 54
\bibitem[Cholis \& Hooper(2013)]{cholis13} Cholis, I., \& Hooper,D.\ 2013, JCAP, 6, 030
\bibitem[Dermer et al. (2014)]{dermer14} Dermer, C. D., Murase, K., Inoue, Y. 2014, arXiv:1406.2633
\bibitem[Guo et al.(2013)]{guo13} Guo, Y.~Q., Hu, H.~B., Yuan, Q., Tian, Z., \& Gao, X.~J.\ 2013, arXiv:1312.7616
\bibitem[Gupta (2013)] {gupta13} Gupta, N.\ 2013, arXiv:1305.4123
\bibitem[He et al.(2013)]{he13} He, H.-N., Wang, T., Fan, Y.-Z., Liu, S.-M., \& Wei, D.-M.\ 2013, \prd, 87, 063011
\bibitem[IceCube Collaboration(2013)] {ic13} IceCube Collaboration\ 2013, Science, 342, 6161
\bibitem[Joshi et al.(2014)]{joshi14} Joshi, J.~C., Winter, W., \& Gupta, N.\ 2014, MNRAS, 439, 3414
\bibitem[Kalashev et al.(2013)]{kalashev13} Kalashev, O.~E., Kusenko, A., \& Essey, W.\ 2013, \prl, 111, 041103
\bibitem[Laha et al.(2013)] {laha13} Laha, R., Beacom, J. F., Dasgupta, B., Horiuchi, S., \& Murase, K.\ 2013, Phys. Rev. D 88, 043009
\bibitem[Liu et al.(2013)]{liu13b} Liu, R.-Y., Wang, X.-Y., Inoue, S., Crocker, R., \& Aharonian, F.\ 2013, arXiv:1310.1263
\bibitem[Liu \& Wang (2013)]{liu13a} Liu, R.-Y., Wang, X.-Y., 2013, \apj, 766, 73
\bibitem[Loeb \& Waxman(2006)]{loeb06} Loeb, A., \& Waxman, E.\ 2006, JCAP, 5, 3
\bibitem[Luis A. Anchordoqui et al.(2013)] {luis13} Luis A. Anchordoqui, Vernon Barger, Ilias Cholis et al.\ 2013, arXiv:1312.6587
\bibitem[Lunardini et al.(2013)]{razzaque13-2} Lunardini, C., Razzaque, S., Theodoseau, K.~T., \& Yang, L.\ 2013, arXiv:1311.7188
\bibitem[M{\"u}cke et al.(2003)]{BLLac1} M{\"u}cke, A., Protheroe, R.~J., Engel, R., Rachen, J.~P., \& Stanev, T.\ 2003, Astroparticle Physics, 18, 593
\bibitem[Murase et al.(2013)]{murase13b} Murase, K., Ahlers, M., \& Lacki, B.~C.\ 2013, \prd, 88, 121301
\bibitem[Murase \& Ioka(2013)]{murase13a} Murase, K., \& Ioka, K.\ 2013, \prl, 111, 121102
\bibitem[Murase et al.(2014)]{murase14} Murase, K., Inoue, Y., \& Dermer, C.~D.\ 2014, arXiv:1403.4089
\bibitem[Neronov et al.(2013)]{neronov13} Neronov, A., Semikoz, D.~V., \& Tchernin, C.\ 2013, arXiv:1307.2158
\bibitem[Padovani \& Resconi(2014)]{BLLac2} Padovani, P., \& Resconi, E.\ 2014, MNRAS, 443, 474
\bibitem[Razzaque (2013)] {razzaque13-1} Razzaque S.\ 2013, arXiv:1310.5123
\bibitem[Roulet et al.(2013)] {roulet13} Roulet, E., Sigl, G., van Vliet, A., \& Mollerach, S. \ 2013, JCAP, 1301, 028
\bibitem[Stecker (2013)]{stecker13} Stecker, F.~W., 2013, \prd 88, 047301
\bibitem[Tamborra et al.(2014)]{tamborra14} Tamborra, I., Ando, S., \& Murase, K.\ 2014, JCAP, 9, 43
\bibitem[Taylor et al.(2014)]{taylor14} Taylor, A.~M., Gabici, S., \& Aharonian, F.\ 2014, arXiv:1403.3206
\bibitem[Wang et al.(2014)]{wang14}Wang, B., Zhao, X. H. \& Li, Z., 2014, JCAP, 11, 028
\bibitem[Waxman \& Bahcall(1997)]{waxman97} Waxman, E., \& Bahcall, J.\ 1997, \prl, 78, 2292
\end{thebibliography}
\end{document}